\documentclass[prb,twocolumn,showpacs]{revtex4}
\usepackage{graphicx}
\usepackage{amsfonts}
\usepackage{amsmath}
\usepackage{amssymb}
\usepackage{bm}
\usepackage{hyperref}
\begin{document}
\title{Kondo-induced hybrid topological insulator in two-dimensional electron system with a quadratic band crossing point}
\author{Rui Wang$^{1,2}$}
\author{Baigeng Wang$^{2}$}
\author{L. Sheng$^{2}$}
\author{D. Y. Xing$^{2}$}
\author{Jian Wang$^{1,3}$}
\affiliation{$^1$Department of Physics and the Center of Theoretical and Computational Physics, The University of Hong Kong, Pokfulam Road, Hong Kong, China\\
$^2$National Laboratory of Solid State Microstructures and Department of
Physics, Nanjing University, Nanjing 210093, China\\
$^3$The University of Hong Kong Shenzhen Institute of Research and Innovation, Shenzhen, P.R. China}
\date{\today }

\begin{abstract}
We investigate the Kondo effect in the two-dimensional electron system with a non-trivial quadratic energy band crossing point. We show that the Kondo effect can induce a new hybrid topological insulator phase which is a coexistence state of the quantum anomalous Hall effect and the TRS-broken quantum spin Hall effect. This hybrid topological insulator exhibits not only a quantized charge Hall current but also a net spin current, which are localized at the edge boundaries. This peculiar topological state arises due to the interplay of two marginally relevant operators, \textit{i.e.}, the Kondo-coupling between the electrons and the local magnetic moment and the electron-electron interaction in the two-dimensional system.
\end{abstract}

\maketitle

\section{introduction}
The most common temperature dependence of the resistivity of metals, $\rho(T)$, either decreases to zero or goes to constant value as $T\rightarrow0$, depending on which  mechanism is dominant, the electron-electron (-phonon) interaction or the non-magnetic disorders. However, a novel resistivity, with $\rho(T)$ being increased as $T\rightarrow0$, is discovered \cite{Haas} in bulk metals containing magnetic impurities. This phenomenon is known as the Kondo effect as it was worked out by Kondo \cite{Kondo} in 1964. The Kondo model describes a single magnetic impurity that is locally coupled to the conduction band electrons. The coupling strength, denoted by $g$, is smeared by the thermal fluctuation and becomes insignificant in the high temperature regime. However, with the temperature being increasingly lowered, it is renormalized to a much larger value. When $T<T_K$ (with $T_K$ being the Kondo temperature), $g$ flows to the strong-coupling limit and leads to the many-body Kondo-singlet state, where the local magnetic moment is completely screened by the conduction band electrons \cite{Hewson}. Here $T<T_K$ is termed as the strong-coupling regime. In this regime, the system shows the Fermi-liquid-like behaviors \cite{Nozi}. Besides, the perturbative theory in terms of $g$ cannot be performed. This difficulty has motivated many exact non-perturbative approaches, including the numerical renormalization group \cite{Wilson,Bulla,Weich}, the Bethe ansatz \cite{Andrei,Gogolin,Konik} and the conformal field theory\cite{Affleck,Ludwig}. On the other hand, in the weak-coupling regime where the energy scale is much larger than the Kondo temperature, a perturbative renormalization group (RG) theory was proposed \cite{Anderson}, where the bare coupling constant is dressed with the RG flow. Through this method, one can obtain a clear understanding of the Kondo physics as well as a satisfactory estimation of the Kondo temperature. Moreover, even though the perturbative RG is in principle applicable only in the weak-coupling regime, it is also very useful to predict possible instabilities at the strong to intermediate couplings. These predictions have received great success in one-dimensional and two-dimensional models in different fields \cite{Dzyaloshinskii,Schulz,Lederer,Song}. The RG method has also been numerically extended to search for instabilities including different topological states of matters \cite{Raghu}.

Recently, with the rapid development of the topological state of matters in condensed matter physics, the Kondo problem is also updated with new perspectives. For example, the Kondo effect in graphene was extensively studied \cite{Novoselov,Geim,Neto,Sarma}, which bears qualitatively different phenomena compared to those in the conventional metals due to the vanishing density of states at the Dirac point \cite{Fritz}. The Kondo problem in the Dirac or Weyl semimetal was also investigated, where the non-Fermi liquid behavior is suggested \cite{Principi}. More remarkably, a new type of topological insulator, termed the topological Kondo insulator \cite{Dzero}, was proposed in the mixed valence compound $\mathrm{SmB}_6$ \cite{Coleman,Alexandrov,Lu}, where the band inversion between the $5d$ and $4f$ band around the X points in the Brillouin
Zone is the main reason for the non-trivial topology. The topological Kondo insulator is classified by the $Z_2$ topological invariant, and is therefore only another realization of the familiar $Z_2$ strong topological insulator \cite{Liang}. As such, it is much more interesting to ask the question: whether any completely new types of topological state of matters (rather than only a realization of the familiar topological states) can be generated by the Kondo-related physics and what is the underlying mechanism?

In this work, we show that a new hybrid topological insulator can be realized in the magnetically doped 2D electron system with a non-trivial quadratic band crossing point (QBCP).
The non-trivial QBCP with $2\pi$ Berry phase has been theoretically predicted to occur in the surface state of the crystalline topological insulator (CTI) \cite{Fu}, as well as in the two dimensional photonic crystal \cite{Chong}. Besides, it is also found in two-dimensional tight-binding electron models with different lattice structures, including the checkerboard \cite{Uebelacker}, Kagome and honeycomb \cite{Wenj} lattices. We study the effect of a Kondo impurity on the QBCP.
Specifically, the electron-electron interaction and the Kondo coupling $g$ are investigated on the equal footing. Using the perturbative RG method, we study the renormalization of the coupling constants, where it is found that both the interactions and the Kondo coupling are marginally relevant in RG sense, and their competition leads to a particularly interesting RG flow of the parameters. In order to determine the possible ground state, a RG-based mean-field theory is constructed to study this strong-coupling regime, where we arrive at several conclusions. (A). In contrast to the conventional Kondo problem, a threshold $g_c$ emerges due to the relevant interactions between electrons.  For $g<g_c$, the screening of the local magnetic moment is suppressed, and the leading instability is the quantum anomalous Hall (QAH) effect. (B) For $g>g_c$, the Kondo-singlet state is developed in the bulk, and it is found to be compatible with the edge states generated by the spontaneous symmetry breaking. (C) For $g>g_c$, the Kondo-effect induces a particular topological phase, which is a coexistence state of the QAH and the quantum spin Hall (QSH) effect. This state possesses both nonzero Chern number and the spin Chern number and enjoys a nonzero quantized charge current and a net transport of spin Hall current that are localized at the edges of the CTI surface. This hybrid topological insulator is a direct consequence of the interplay between two marginally relevant operators, \textit{i.e.}, the electron-electron interaction and the Kondo-coupling.

\section{Model and Hamiltonian}
We consider a two dimensional electron system with a QBCP, which can be described by the Hamiltonian \cite{Fu}, $H=\sum_{\mathbf{k}}\Psi^{\dagger}({\mathbf{k}})\mathcal{H}_0(\mathbf{k})\Psi(\mathbf{k})$, where $\Psi({\mathbf{k}})=[\psi_{\uparrow,1}({\mathbf{k}}),\psi_{\downarrow,1}({\mathbf{k}}),\psi_{\uparrow,2}({\mathbf{k}}),\psi_{\downarrow,2}({\mathbf{k}})]^T$, with $\psi_{s,\alpha}({\mathbf{k}})$ being the annihilation operator for the electron with the spin $s$ and the flavor $\alpha$. $\alpha$ represent the pseudo spin degree of freedom, whose physical meaning depends on the specific models. It can either denote the sublattice or the orbital degrees of freedom. In what follows, we regard $\alpha$ as the orbital index. Then, the single particle Hamiltonian of the QBCP reads,
\begin{equation}\label{eq1}
  \mathcal{H}_0(\mathbf{k})=(d_I(\mathbf{k})I_{\sigma}+d_x(\mathbf{k})\sigma^x+d_z(\mathbf{k})\sigma^z)\tau^0,
\end{equation}
with
\begin{eqnarray}
  d_I(\mathbf{k}) &=& t_I(k_x^2+k_y^2), \\
  d_x(\mathbf{k}) &=& 2tk_xk_y, \\
  d_z(\mathbf{k}) &=& t(k^2_x-k^2_y),
\end{eqnarray}
where $\sigma$ and $\tau$ denote the orbital and spin degrees of freedom respectively and $I_{\sigma}$ is the identity matrix in the orbital space. In the following, we focus on the particle-hole symmetric case where  $t_I=0$. The special feature of the above Hamiltonian, as was studied in Ref.\cite{Sun}, is that it depicts a QBCP with the Berry phase $2\pi$. This type of energy node is robust in the sense that the lattice symmetry ($\mathrm{C}_4$ or $\mathrm{C}_6$) or the time-reversal symmetry (TRS) has to be broken in order to lift the energy degeneracy and open up a gap, leading to the split of the QBCP into Dirac cones.  However, even though QBCP is protected by these symmetries, it shows a fragility against electron-electron interactions since instabilities will develop for any non-vanishing interaction, resulting in the spontaneous breaking of the lattice symmetry or the TRS. Different from the general case in Ref.\cite{Sun}, the QBCP considered here enjoys both the orbital and spin flavor. The orbital and spin degrees of freedom complicate the interactions by bringing about more channels, such as the inter- and intra-orbital couplings. In the following, we study the case where the screening effect is significant and focus on the short-ranged interactions between electrons, as described by the following local model
\begin{eqnarray}
  V_1 &=& v_1\sum_{\{\mathbf{k}\}}\psi^{\dagger}_{s,1}(\mathbf{k}_1)\psi^{\dagger}_{s^{\prime},2}(\mathbf{k}_2)\psi_{s^{\prime},2}(\mathbf{k}_3)\psi_{s,1}(\mathbf{k}_4), \\
  V_3 &=& v_3\sum_{\{\mathbf{k}\}}\psi^{\dagger}_{s,2}(\mathbf{k}_1)\psi^{\dagger}_{s^{\prime},1}(\mathbf{k}_2)\psi_{s^{\prime},2}(\mathbf{k}_3)\psi_{s,1}(\mathbf{k}_4),
\end{eqnarray}
with $V_1$ and $V_3$ being the intra-orbital and the inter-orbital scattering, respectively. The sum, $\sum_{\{\mathbf{k}\}}$, denotes  ``$\sum_{\mathbf{k}_1,\mathbf{k}_1,\mathbf{k}_1,\mathbf{k}_1}\delta(\mathbf{k}_1+\mathbf{k}_2-\mathbf{k}_3-\mathbf{k}_4)$" and the sum for repeated index is implied. In what follows, we consider the repulsive couplings with $v_i\geq0$ ($i=1,3$).
Moreover, in order to extract the physics clearly, we have temporarily neglected the other two interactions $V_2$ and $V_4$, which describe the scattering between electrons within the same orbital and the coupling process where two electrons are both scattered from one orbital to the other, respectively. The effect of these two more couplings will be calculated and discussed in detail in section \uppercase\expandafter{\romannumeral6}.

Besides the interactions, we are interested in the effect of a single magnetic impurity on the QBCP, which can be investigated using the Kondo model. By adopting the pseudo fermion representation of the local magnetic moment, $\mathbf{S}=\frac{1}{2}f^{\dagger}_{\sigma}\mathbf{\sigma}_{\sigma,\sigma^{\prime}}f_{\sigma^{\prime}}$ with the constraint $\sum_{\sigma}f^{\dagger}_{\sigma}f_{\sigma}=1$ and $f_{\sigma}$ being the annihilation operator for the pseudo fermion, the interaction between the local magnetic moment and the conduction electron is written as
\begin{equation}\label{eq2}
  H_K=g\sum_{\mathbf{k},\mathbf{k}^{\prime}}\psi^{\dagger}_{s1}(\mathbf{k})\psi_{s^{\prime},1}(\mathbf{k}^{\prime})f^{\dagger}_{s^{\prime}}f_{s},
\end{equation}
where $g$ is the s-d coupling constant. Here we have assumed that the local moment is only coupled to one orbital. The coupling to both the two orbitals constitutes a two-channel Kondo problem, which we left for future investigation. Eq.(1)-(7) constitutes the main model we are interested in this work. Since this model is essentially a combined problem of the Kondo effect and the QBCP, we term it by the Kondo-QBCP (KQBCP) model for the purpose of brevity.

As is well known, for anti-ferromagnetic coupling $g>0$ (which we consider), the RG analysis on the traditional Kondo problem in metals shows two fixed points, $g=0$ and $g=+\infty$. For a nonzero bare $g$, the renormalized coupling gets stronger and stronger when the temperature is lowered, and finally leads to the complete screening of the local magnetic moment, forming the well-known Kondo singlet state. However, in the current case, due to the fragility of the QBCP, the interaction between electrons is also a marginally relevant operator, which may lead to various ground states with symmetry spontaneously broken. So, there exists two trends towards two different strong coupling limits in this model. The questions naturally arise as whether the two trends cooperate or compete with each other, and what is the most relevant order formed at low temperature. In this paper, we address these questions using the combined method of the perturbative RG and the mean-field theory.
\section{Renormalization group flow of the coupling constants}
Resorting to the functional path integral representation of the partition function $Z=\int D\overline{\Psi}D\Psi D\overline{f}Df e^{S}$ \cite{Shankar}, and introducing a Lagrangian multiplier \cite{Principi} to enforce the number constraint condition of the pseudo fermions, we arrive at the imaginary-time action $S$ describing the KQBCP model,
\begin{equation}\label{eq3}
\begin{split}
  S&=\int d\tau d\mathbf{k}[\Psi^{\dagger}(-\partial_{\tau}-\mathcal{H}_0)\Psi+f^{\dagger}_s(-\partial_{\tau}-\lambda)f_s]\\
  &-v_1\int d\tau \prod^4_id^{\prime}\mathbf{k}_i\psi^{\dagger}_{s,1}(\mathbf{k}_1)\psi^{\dagger}_{s^{\prime},2}(\mathbf{k}_2)\psi_{s^{\prime},2}(\mathbf{k}_3)\psi_{s,1}(\mathbf{k}_4)\\
  &-v_3\int d\tau \prod^4_id^{\prime}\mathbf{k}_i\psi^{\dagger}_{s,2}(\mathbf{k}_1)\psi^{\dagger}_{s^{\prime},1}(\mathbf{k}_2)\psi_{s^{\prime},2}(\mathbf{k}_3)\psi_{s,1}(\mathbf{k}_4)\\
  &-g\int d\tau d\mathbf{k}d\mathbf{k}^{\prime}\psi^{\dagger}_{s,1}(\mathbf{k})\psi_{s^{\prime},1}(\mathbf{k}^{\prime})f^{\dagger}_{s^{\prime}}f_{s}.
\end{split}
\end{equation}
For brevity, the dependence of $\psi$ on $\tau$ is implicit  and the $\delta$ function enforcing the momentum conservation is not explicitly written but represented by ``$^\prime$" in the integral. We shall focus on the most interesting case where the Fermi energy lies at the QBCP by setting the chemical potential of the electrons to zero\cite{rui}. As one can see in the above equation, the Lagrangian $\lambda$ acts as the chemical potential of the pseudo fermions, which, as will be shown later, should be calculated self-consistently so that the pseudo fermion number condition is satisfied at the mean-field level. Since the QBCP is the low-energy effective description of the tight-binding models \cite{Wenj,Uebelacker}, a cutoff $\Lambda_0$ is implied.
As what RG does, one can decrease $\Lambda_0$ step by step to obtain the effective action with renormalized parameters for the coarse-grained model, integrating out the degree of freedom with the larger momentum.

Through the standard RG analysis, one can find that the three coupling constants, $v_1$, $v_3$ and $g$ are all marginal at tree level \cite{Shankar,Sun}, therefore the perturbative RG to one-loop order needs to be performed. In order to do so, all the topologically distinct one-loop diagrams should be calculated, which can be classified into three different channels, i.e., the $\mathrm{ZS}$, $\mathrm{ZS}^{\prime}$ and the $\mathrm{BCS}$ diagrams \cite{Shankar}. Moreover, a key point lies in the choice of basis. Here, we perform the RG calculation in the $\Psi({\mathbf{k}})$ basis (see section \uppercase\expandafter{\romannumeral2}), rather than the diagonalized basis of $\mathcal{H}_0$. This is important due to the following reasons. Transforming $\mathcal{H}_0$ to its diagonalized basis leads to the orbital make-ups in the interactions \cite{Uebelacker}, which should not be neglected in order to keep the complete topological behavior of the QBCP. However, in RG sense, the high order expansion of $\mathbf{k}$ of the orbital make-ups are irrelevant, only leaving a few constant bare couplings that are marginal at tree level \cite{Shankar,Song}. Therefore, the perturbative RG treatment in the diagonalized basis can only present incomplete predictions of the leading instability. In comparison, even though it is much more complicated, the $\Psi({\mathbf{k}})$ basis can avoid this problem. In this basis, the Green's functions of the non-interacting electrons are matrices, whose elements are $G_{s,s^{\prime};\alpha,\beta}=G_{s,s;\alpha,\beta}\delta_{s,s^{\prime}}$, with $G_{\uparrow,\uparrow;\alpha,\beta}=G_{\downarrow,\downarrow;\alpha,\beta}=G_{\alpha,\beta}$. So, when treating the second order perturbation, one would encounter averages over the fast mode of the loop-momentum both in the particle-particle and in the particle-hole channel with different orbital flavors. We represent these averages by loops in Feynman diagram. Examples in both the particle-hole and the particle-particle channels are shown in Fig.1, where the dashed lines represent the fast modes. Performing Gaussian integrals in the diagrams, we obtain the results of all the averages (over fast mode) needed in the RG calculation, which are listed below,
\begin{figure}[tbp]
\includegraphics[width=3.4in]{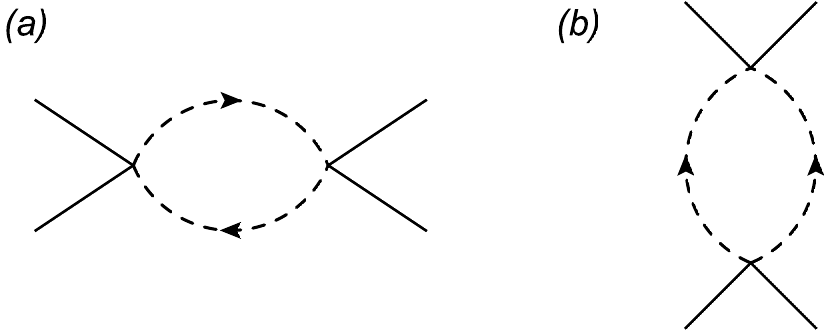}
\caption{(color online) Typical diagrams in the one loop order RG calculations. (a) denotes the average in the particle-hole channel and (b) represents the average in the particle-particle channel. Different vertices $v_1$, $v_3$ and $g$ bring about various notations in terms of spin and orbital degrees of freedom, which is not shown for the purpose of brevity.}
\end{figure}
\begin{eqnarray}
   <G_{11}G_{22}>^{p-h}  &=& -\frac{3}{16\pi t}dl,\\
  <G_{11}G_{22}>^{p-p} &=& \frac{1}{16\pi t}dl, \\
  <G_{12}G_{12}>^{p-h} &=& <G_{12}G_{12}>^{p-p}=\frac{1}{16\pi t}dl, \\
  <G_{11}G_{11}>^{p-h} &=& <G_{22}G_{22}>^{p-h}=-\frac{1}{16\pi t}dl, \\
  <G_{11}G_{11}>^{p-p} &=& <G_{22}G_{22}>^{p-p}=\frac{3}{16\pi t}dl, \\
  <G_{12}G_{11}>^{p-h} &=& <G_{12}G_{22}>^{p-h}=0,
\end{eqnarray}
where $dl=\ln{\Lambda/\Lambda^{\prime}}$ is the RG flowing parameter, with $\Lambda$ being the cutoff at a generic RG step and $\Lambda^{\prime}$ being the reduced cutoff for the coarse-grained Hamiltonian at the next RG step. Then following the standard RG procedure, we obtain the coupled RG flow rate equations of the coupling constants $v_1$, $v_3$ and $g$ under the one-loop correction,
\begin{eqnarray}
  \frac{dv_1}{dl} &=&\frac{1}{4\pi}(v^2_1-\frac{1}{2}v_1v_3-\frac{1}{4}v^2_3), \\
  \frac{dv_3}{dl} &=& -\frac{1}{2\pi}(v^2_3-\frac{1}{4}v_1v_3+\frac{1}{8}v^2_1), \\
  \frac{dg}{dl} &=& \frac{1}{2\pi}(g^2-\frac{1}{8}v_3g),
\end{eqnarray}
where we have rescaled the interactions by $v_{1,3}\rightarrow v_{1,3}/t$ and $g\rightarrow g/t$. As is seen from above equations, only the inter-orbital interaction $V_3$ explicitly modifies $g$. Even though $V_1$ does not dress the flow of $g$ by itself, it is also essential since it will affect $g$ via the indirect modification on $V_3$. The running of the coupling constants leads to the fixed points in the three-dimensional parameter space. Several observations can be drawn from the above equation. First, when both the bare interactions $v^0_1$, $v^0_3$, and the bare Kondo coupling $g^0$ are zero, the system always lies at the fixed point, $(v_1,v_3,g)=(0,0,0)$, which stands for the initial non-interacting QBCP. Second, when the interaction is absent but the bare Kondo coupling $g^0$ is nonzero, the flow equation is reduced to that of the traditional Kondo problem, where $g$ is renormalized to larger and larger value and arrives at the fixed point $(v_1,v_3,g)=(0,0,+\infty)$, which represents the formation of the Kondo-singlet state. Despite the above observation, however, our real interest lies in the situation where both the bare interaction and the bare Kondo-coupling are nonzero. In this case, solving the flow equations numerically, we find an interesting renormalization flow of the interactions, which is plotted in Fig.2. As is shown, for generic bare interactions (except for $v^0_3\gg v^0_1$), $v_1$ and $v_3$ always tend to diverge toward large positive and negative values respectively when the energy scale is gradually lowered, \textit{i.e.}, $v_1\rightarrow v$ and $v_3\rightarrow -v$, with $v\gg v^0_{1,3}$. This renormalization behavior is very interesting since it will lead to a peculiar topological state when the magnetic moment is present (see below), and therefore is our main focus in this work. In plotting Fig.2, we do not show the regime where $v_{1,3}$ diverge since this regime is beyond the perturbative treatment and the RG flow fails to give an exact description.
\begin{figure}[tbp]
\includegraphics[width=3.4in]{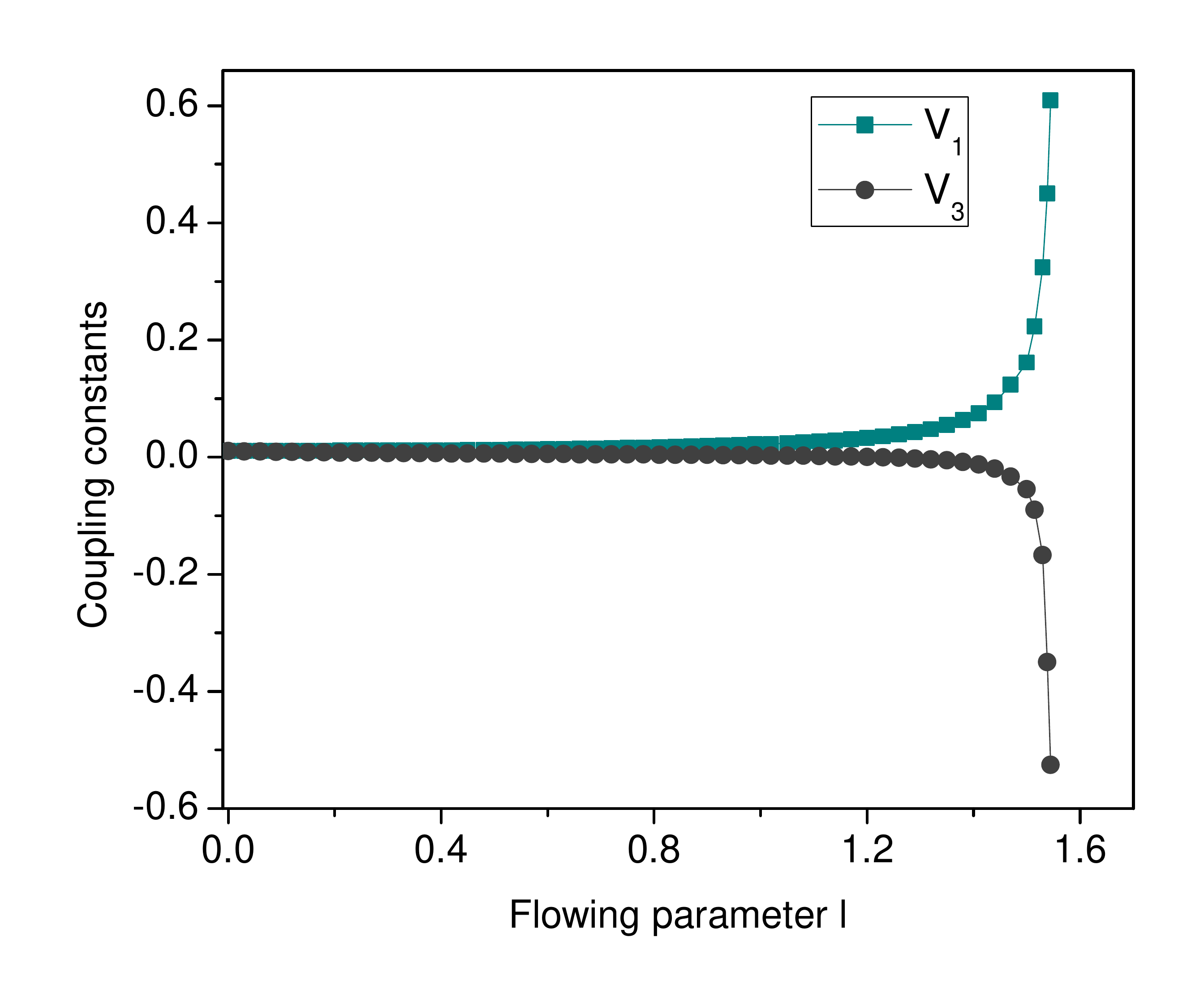}
\caption{(color online) The RG-flow of the coupling constants $v_1$, $v_3$ versus the flowing parameter $l$. The green and grey curves represent for the couplings $v_1$, $v_3$, respectively. The bare coupling constants are chosen as $v^0_1=v^0_3=0.01$. Similar flows can always be found despite the choice of the initial bare parameters.}
\end{figure}
\section{Susceptibilities}
Based on the flow equation of the coupling constants, we can further determine what is the most relevant instability of the KQBCP model. To do so, we introduce test susceptibilities, \textit{i.e.}, the quadratic perturbations, into the action Eq.\eqref{eq3}, and then investigate their RG flow up to one-loop order. The earliest divergent susceptibility will indicate the most possible ground state at the low temperature \cite{Song}.

First, following Ref.\cite{Sun}, we list all the possible vertices or the marginal operators in the Table \uppercase\expandafter{\romannumeral1}, including the (spontaneous) rotation symmetry-breaking nematic phases, \textit{i.e.}, the nematic semimetal \uppercase\expandafter{\romannumeral1} (NS\uppercase\expandafter{\romannumeral1}), nematic semimetal \uppercase\expandafter{\romannumeral2} (NS\uppercase\expandafter{\romannumeral2}), nematic-spin-nematic \uppercase\expandafter{\romannumeral1} (NSN\uppercase\expandafter{\romannumeral1}), and the nematic-spin-nematic \uppercase\expandafter{\romannumeral2} (NSN\uppercase\expandafter{\romannumeral2}), as well as the (spontaneous) TRS-breaking QAH and its triplet counterpart, the QSH state, and the magnetic ordering (MO) phase. Then, considering the one-loop RG correction to these introduced vertices, we obtain the renormalization of the coupling constants using the same method as before. The results reads,
\begin{table}[b]
\caption{\label{tab:table1}
All possible susceptibilities as the test vertices
}
\begin{ruledtabular}
\begin{tabular}{ccc}
Test vertices & Phases & $\Gamma_j$\\
\colrule
$\Phi\sum_{\mathbf{k}}\Psi^{\dagger}(\mathbf{k})\sigma^y\Psi(\mathbf{k})$ & QAH & $\frac{1}{4\pi}(v_1-2v_3)$\\
$Q_1\sum_{\mathbf{k}}\Psi^{\dagger}(\mathbf{k})\sigma^z\Psi(\mathbf{k})$ &  NS\uppercase\expandafter{\romannumeral1} & $\frac{1}{4\pi}(v_1-v_2-\frac{1}{2}v_3)$ \\
$Q_2\sum_{\mathbf{k}}\Psi^{\dagger}(\mathbf{k})\sigma^x\Psi(\mathbf{k})$ &  NS\uppercase\expandafter{\romannumeral2} & $\frac{1}{8\pi}(v_1-2v_3)$ \\
$\mathbf{Q}^t_1\sum_{\mathbf{k}}\Psi^{\dagger}(\mathbf{k})\overrightarrow{\tau}\sigma^z\Psi(\mathbf{k})$ & NSN\uppercase\expandafter{\romannumeral1} & $\frac{1}{4\pi}(v_1-v_2-\frac{1}{2}v_3)$\\
$\mathbf{Q}^t_2\sum_{\mathbf{k}}\Psi^{\dagger}(\mathbf{k})\overrightarrow{\tau}\sigma^x\Psi(\mathbf{k})$ &  NSN\uppercase\expandafter{\romannumeral2} & $\frac{1}{8\pi}(v_1-2v_3)$\\
$\mathbf{\Phi}^t\sum_{\mathbf{k}}\Psi^{\dagger}(\mathbf{k})\overrightarrow{\tau}\sigma^y\Psi(\mathbf{k})$ & QSH & $\frac{1}{4\pi}(v_1-2v_3)$\\
$\Delta_K\sum_{\mathbf{k}}\psi^{\dagger}_{s,1}f_s$ & Kondo-singlet & $\frac{g}{2\pi}$\\
$\mathbf{m}\sum_{\mathbf{k}}\Psi^{\dagger}(\mathbf{k})\overrightarrow{\tau}\sigma^0\Psi(\mathbf{k})$ & MO & 0
\end{tabular}
\end{ruledtabular}
\end{table}
\begin{equation}
\Delta^{\prime}_j=\Delta_j(1+\Gamma_jdl),
\end{equation}
where $j$ denotes the different phases in Table. \uppercase\expandafter{\romannumeral1}, $\Delta_j$ and $\Delta^{\prime}_j$ represent the corresponding bare and dressed order parameters respectively. $\Gamma_j$ is the calculated susceptibilities, which have been listed in the third column of Table. \uppercase\expandafter{\romannumeral1}.
In the RG sense, the divergence of the susceptibility $\Gamma_j$ implies a phase transition from the free QBCP to these various spontaneous symmetry-breaking orders in Table \uppercase\expandafter{\romannumeral1}.
\begin{figure}[tbp]
\includegraphics[width=3.4in]{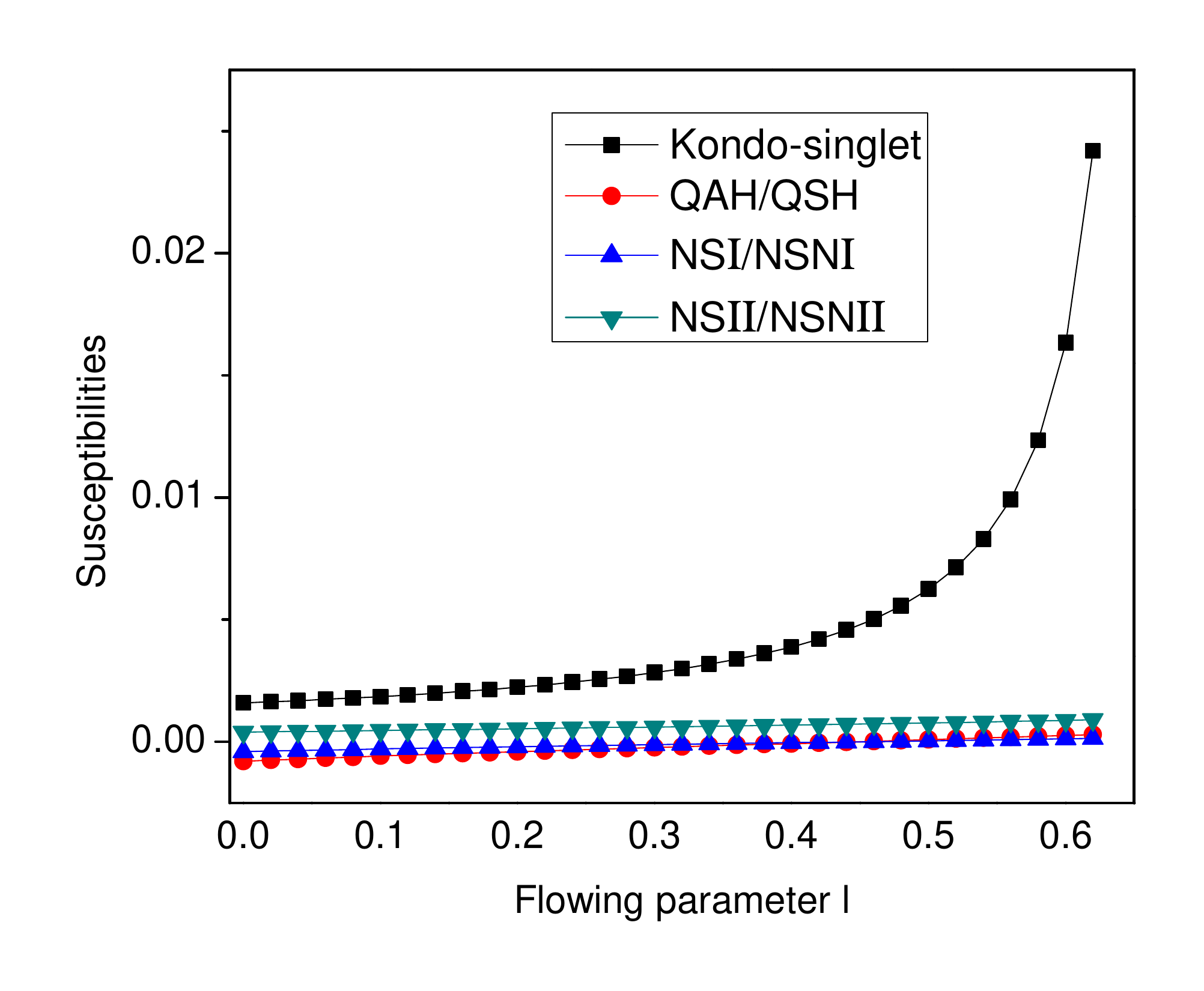}
\caption{(color online) The flow of the susceptibilities $\Gamma_j$ versus the flowing parameter $l$. The black curve denotes the $\Gamma$ of the Kondo-singlet state, which is the leading instability. The bare parameters are chosen as $g^0=v^0_1=v^0_3$}
\end{figure}
The susceptibilities for the Kondo-singlet state is plotted by the black curve in Fig.3 for bare couplings $g^0=0.1$ and $v^0_1=v^0_3=0.1$.  It shows that the Kondo-singlet emerges as the leading instability.  In this case where Kondo-singlet state is developed, we have further investigated the effect of $v_1$, $v_3$ on the Kondo temperature $T_K$. It is found that, for repulsive interactions, $T_K$ is much more dependent on $v^0_3$ than $v^0_1$, and it decreases with the increase of $v^0_3$. This means that the interactions hampers the formation of the Kondo-singlet state, which can also be obviously seen from the flow equations Eq.(15)-(17).
In Fig.4, we search for the leading instabilities besides the Kondo-singlet state. Two conclusions can be drawn from Fig.4. First, the triplet orderings have the same susceptibilities as their singlet counterparts, and they cannot be distinguished by the RG-flow alone, which is consistent with the conclusion drawn by Ref.\cite{Sun}. Second, as is shown, the susceptibility for the QAH or the QSH states are always the first one to get divergent, suggesting the QAH or QSH to be the leading instabilities, instead of the NS\uppercase\expandafter{\romannumeral1}/NSN\uppercase\expandafter{\romannumeral1}, and the NS\uppercase\expandafter{\romannumeral2}/NSN\uppercase\expandafter{\romannumeral2} phases.
\begin{figure}[tbp]
\includegraphics[width=3.4in]{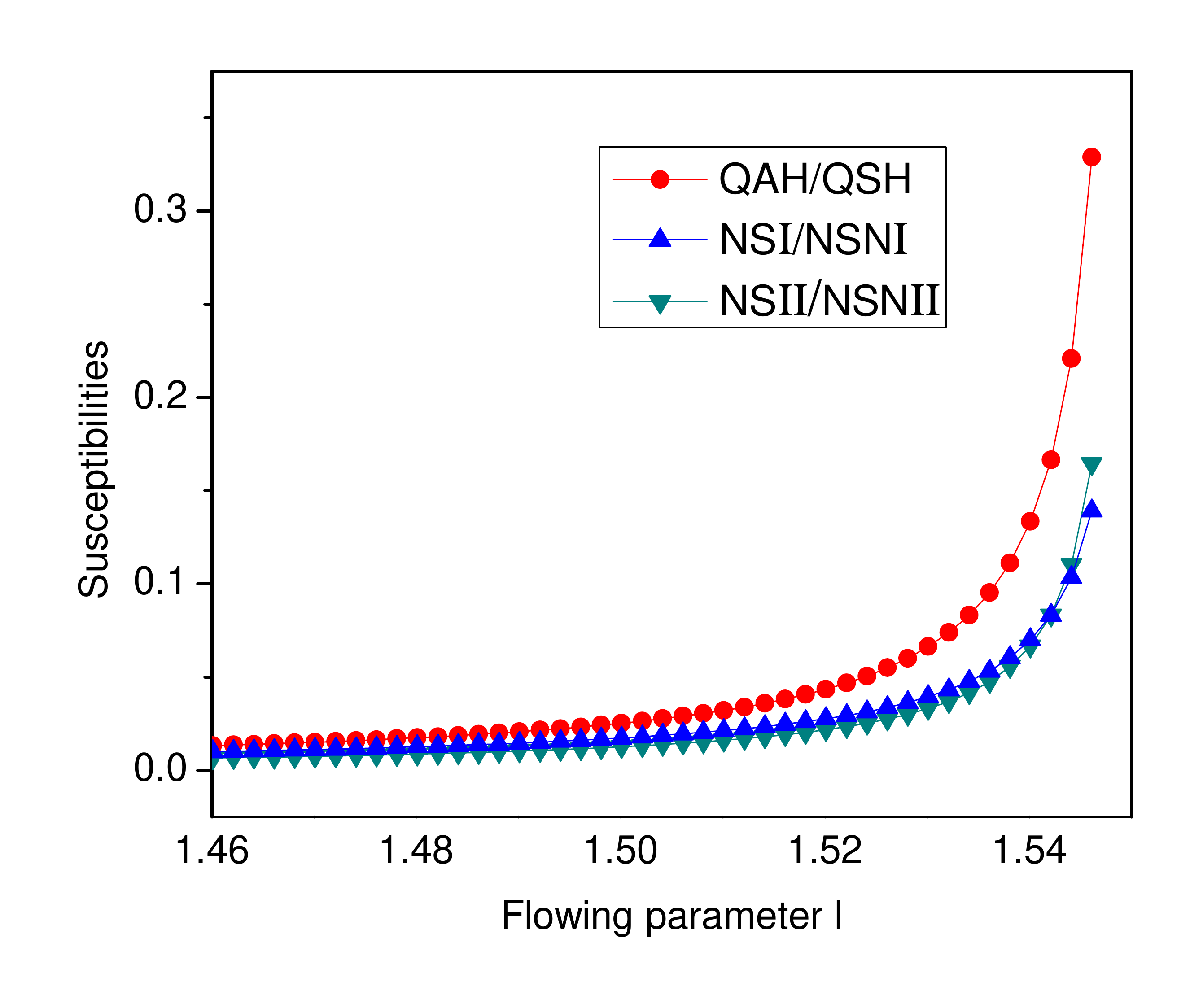}
\caption{(color online) The flow of the susceptibilities $\Gamma_j$ other than the Kondo-singlet state. The red curve denotes the susceptibilities of the QAH and the QSH states (which are the same), the blue curve represents for susceptibilities of the NS\uppercase\expandafter{\romannumeral1} and the NSN\uppercase\expandafter{\romannumeral1} states, and the green curve describes the susceptibilities of the NS\uppercase\expandafter{\romannumeral2} and the NSN\uppercase\expandafter{\romannumeral2} phases. The bare parameters are chosen as $g^0=v^0_1=v^0_3$}
\end{figure}
The above observation provides us with both insights and puzzles in understanding the true ground state in the KQBCP model. On one hand, insights are obtained as: (A). The complete screening of the local magnetic moment and the formation of the Kondo-singlet are still likely to take place (as long as $g^0$ is large enough) even though the interaction is present, as is shown in Fig.3. (B). The interaction can lower the Kondo temperature and make the Kondo-singlet more difficult to occur.  (C). Besides the Kondo-singlet, the leading instability is the QAH and QSH states.  On the other hand, problems still remain as following. (A). Can the Kondo-singlet coexist with the symmetry-breaking phases? (B). Which order is more likely to take place, the QAH or the QSH? This cannot be answered by the above RG analysis since the RG method is unable to extract the energetics of the relevant states, which is necessary to determine which phases are more stable. Therefore, in what follows, we solve these problems by taking one step further, using a RG-based mean-field theory.

\section{Mean-field theory and the Hybrid topological insulator}
From Sec.\uppercase\expandafter{\romannumeral3} and Sec.\uppercase\expandafter{\romannumeral4}, two important conclusions are obtained. (A). Besides the local Kondo-singlet, the leading orders are found to be the QAH or the QSH states.  (B). When both the electron-electron interaction and the Kondo coupling are present in the system, the interactions flow to the strong-coupling regime, where $v_1$ and $v_3$ tends to move toward large positive and negative values respectively. Based on the two conclusions, a mean-field theory can be performed by studying the dressed Hamiltonian at certain low energy scale, which is obtained by substituting the bare coupling constants by the renormalized ones. In the mean-field level, we introduce the bosonic mean-fields in the channels of the Kondo-singlet, QAH and the QSH orderings, \textit{i.e.}, $\Phi=<\Psi^{\dagger}(\mathbf{r})\sigma^y\tau^0\Psi(\mathbf{r})>$, $\mathbf{\Phi}^t=<\Psi^{\dagger}(\mathbf{r})\sigma^y\overrightarrow{\tau}\Psi(\mathbf{r})>$ and  $\Delta_K=g<\sum_{\mathbf{k}}\psi^{\dagger}_{s,1}(\mathbf{k})f_s>$, and make Hubbard-Stratonovich decomposition of the interactions.  This treatment leads to the following Hamiltonian,
\begin{equation}\label{eq4}
\begin{split}
  H_{MF}&=\sum_{\mathbf{k}}\Psi^{\dagger}(\mathbf{k})[\mathcal{H}_0(\mathbf{k})-\frac{3v}{2}\sigma^y\tau^0\Phi-\frac{v}{2}\sigma^y\mathbf{\Phi}^t\cdot\overrightarrow{\tau}]\Psi(\mathbf{k})\\
&-g\Delta_K\sum_{\mathbf{k},s}f^{\dagger}_s\psi_{s,1}(\mathbf{k})-g\Delta^{\star}_K\sum_{\mathbf{k},s}\psi^{\dagger}_{s,1}(\mathbf{k})f_s\\
&+\frac{v}{4}\sum_{\mathbf{k}}(3\Phi^2+\mathbf{\Phi}^{t2}+\frac{4g\Delta^2_K}{v})+\frac{\lambda}{N}\sum_{\mathbf{k},s}(f^{\dagger}_{s}f_{s}-1).
\end{split}
\end{equation}
Through a Bogoliubov transformation in the mixing basis of the electron and the pseudo fermion \cite{Perkins}, the energy spectrum $E_n(\mathbf{k})$ can be obtained, with $n$ denoting the band index. Then the self-consistent equations are obtained by  minimizing the mean-field energy, i.e., $\sum_{\mathbf{k},n}\partial E_n(\mathbf{k})/\partial C=0$, where $C$ represents the mean-field parameters $\Delta_K$, $\Phi$, $\mathbf{\Phi}^t$ and the Lagrangian multiplier $\lambda$. The self-consistent equations can be numerically solved and the saddle points can be extracted for different parameters $v$ and $g$, the result of which is shown in Fig.5. Three conclusions can be drawn from Fig.5. First, with $v$ fixed, we find that $\Delta_K=0$ for small $g$, while $\Delta_K\neq0$ for large $g$. So, compared with the traditional Kondo problem where no QBCP is present, there emerges a threshold $g_c$ above which the Kondo-singlet will be developed. This is the result of the renormalization of $v_3$ on $g$ (see Eq.(17)). Second, since the mean-field order parameters undergo abrupt stepwise changes at $g_c$, we conclude that a quantum phase transition takes place at this critical point. As shown in Fig.5, in the region $g<g_c$, the nonzero $\Phi$ indicates a simple QAH phase. In the region $g>g_c$, both QAH and QSH phases develop nonzero order parameters. Therefore, we observe the coexistence of the QAH and the QSH states, forming a new hybrid topological state. This new hybrid topological insulator only emerges when the Kondo coupling $g$ reaches the threshold $g_c$, and its formation is due to the interplay between the electron-electron interaction and the s-d coupling.
\begin{figure}[tbp]
\includegraphics[width=3.6in]{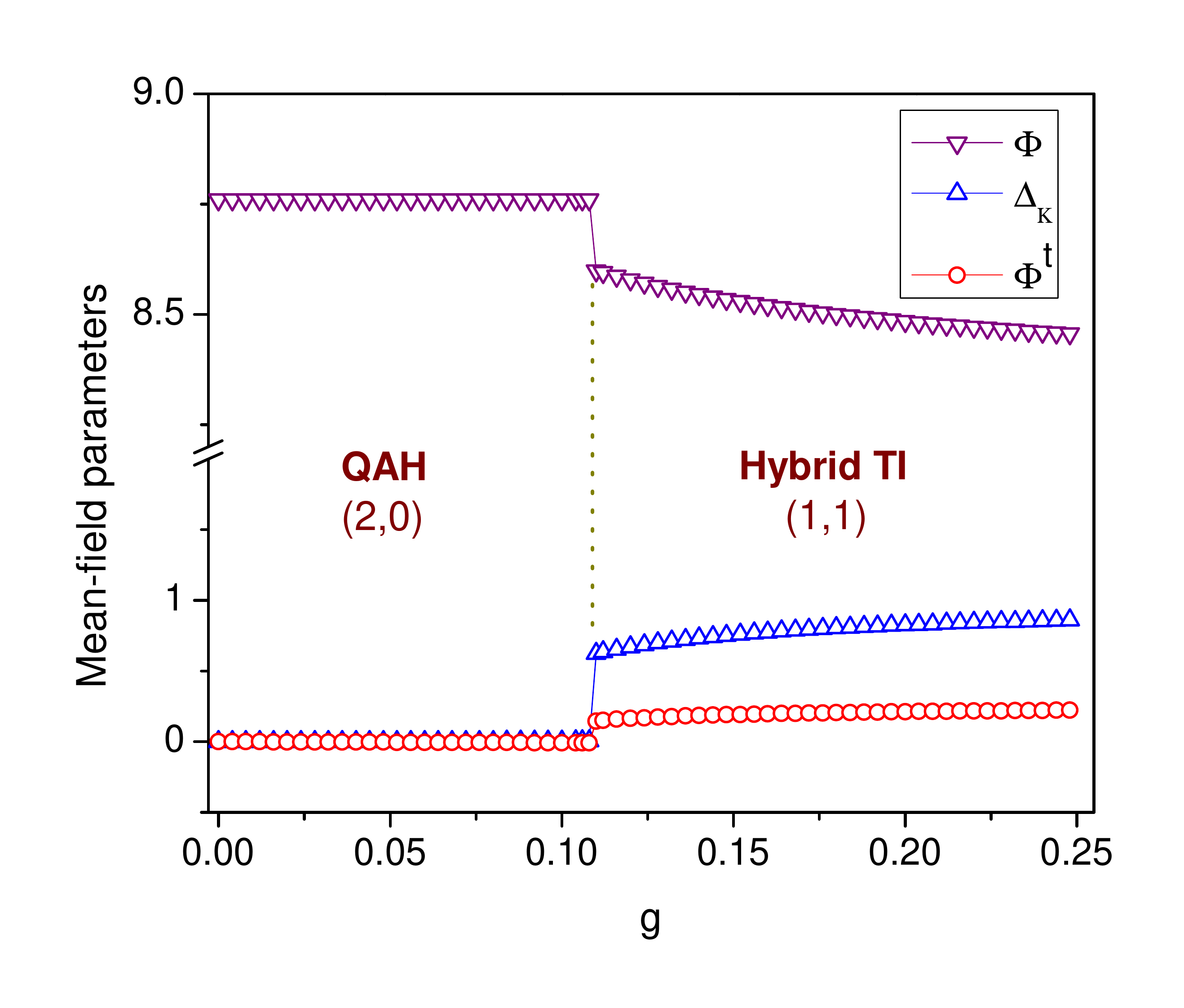}
\caption{(color online) The self-consistent solution of the mean-field order parameters, $\Phi$, $\mathbf{\Phi}^t$ and $\Delta_K$, represented by the purple, red and blue curves, respectively. The calculated Chern number and spin Chern number $(C,C_S)$ are shown to distinguish different phases. The parameters chosen are $v=0.8$. similar results are valid for other values of parameters.}
\end{figure}

In order to understand the corresponding phases, we investigate the nontrivial topology of the Hamiltonian Eq.\eqref{eq4} by calculating the Chern number $C$ and the spin Chern $C_S$ number \cite{dnsheng} for $g<g_c$ and $g>g_c$ respectively, which is shown in Fig.5. For $g<g_c$, the Chern numbers are calculated to be $(C,C_S)=(2,0)$. The high Chern number $C=2$ is due to the d-wave characteristic of the QBCP. In this case, the relevant interactions lead to the spontaneous breaking of TRS, developing a mass term $\Phi$ and opening up a gap at the QBCP. Similar to the linear Dirac node, the gapped TRS-broken QBCP also bears a QAH effect, but with the Hall conductance doubled. For $g>g_c$, the Chern number and spin Chern number are obtained as $(C,C_S)=(1,1)$.   Both the Chern numbers and the spin Chern number are nonzero, this verifies the coexistence region where both $\Phi$ and $\mathbf{\Phi}^t$ are nonzero.

The coexistence state found here is closely related to our previous work, which discusses the fate of the QSH effect with the presence of a TRS-breaking Zeeman field \cite{Yang}. Using the spin Chern number, we found in Ref.\cite{Yang} that the nontrivial topological properties of QSH systems remain intact even when the TRS is broken \cite{Yang}. The breaking of the TRS leads to a small gap of the QSH edge. Remarkably, even though a small gap is opened up, the TRS-broken QSH is still topologically distinct from the trivial state, \textit{i.e.}, the edge states become quasi-helical and are robust as long as the bulk energy gap is not closed. In this phase, there is usually a weak scattering between forward and backward movers, as evidenced by the small energy gap in the edge state spectrum, leading to a low-dissipation spin transport. The TRS-broken QSH phase has also been found in the $\mathrm{Cr}$-doped $(\mathrm{Bi,Sb})_2\mathrm{Te}_3$\cite{jwang} and in the spin-orbit coupled electron system with staggered magnetic fluxes \cite{yuan}, where the quasi-helical QSH edge state coexists with the chiral QAH edge state on the sample edge.
The hybrid topological insulator phase we predicted here enjoys the same bulk topology with that of Ref.\cite{jwang,yuan}.  The nonzero $C=1$ implies that the 2D electron system has an exactly quantized charge Hall conductance due to the gapless chiral QAH edge states, and the nonzero $C_s=1$ suggests that it exhibits a net transport of spin owing to the quasi-helical QSH edge state \cite{Yang}.


\section{Effects of other interaction channels}
In this section, we take into account the other two local scattering processes, which are described by the following interactions
\begin{eqnarray}
  V_2 &=& v_2\sum_{\{\mathbf{k}\}}\psi^{\dagger}_{s,\alpha}(\mathbf{k}_1)\psi^{\dagger}_{\overline{s},\alpha}(\mathbf{k}_2)\psi_{\overline{s},\alpha}(\mathbf{k}_3)\psi_{s,\alpha}(\mathbf{k}_4),  \\
  V_4 &=& v_4\sum_{\{\mathbf{k}\},\alpha}\psi^{\dagger}_{s,\alpha}(\mathbf{k}_1)\psi^{\dagger}_{\overline{s},\alpha}(\mathbf{k}_2)\psi_{\overline{s},\overline{\alpha}}(\mathbf{k}_3)\psi_{s,\overline{\alpha}}(\mathbf{k}_4).
\end{eqnarray}
Here $V_2$ denotes the coupling between electrons from the same orbital and $V_4$ describes the interaction process where two electrons are scattered from one orbital to the other. The corrections from these two vertices will lead to new renormalization flow of the coupling constants. Following the same perturbative RG approach as before, we obtain the coupled flow equations in the five-dimensional parameter space, which reads
\begin{eqnarray}
  \frac{dv_1}{dl} &=&\frac{1}{4\pi}(v^2_1-2v_1v_2-\frac{1}{2}v_1v_3-\frac{1}{4}v^2_3), \\
  \frac{dv_2}{dl} &=&-\frac{1}{4\pi}(v^2_2+v_2v_3+v_2v_4+\frac{3}{2}v_4^2), \\
  \frac{dv_3}{dl} &=& -\frac{1}{2\pi}(v^2_3-\frac{1}{4}v_1v_3+\frac{1}{8}v^2_1), \\
  \frac{dv_4}{dl} &=&-\frac{1}{4\pi}(v_4^2+\frac{1}{2}v_2^2+3v_2v_4-v_1v_4),\\
  \frac{dg}{dl} &=& \frac{g^2}{2\pi}-\frac{v_3g}{16\pi}.
\end{eqnarray}
The new RG flows describing the renormalization of the interactions for small values of $v^0_2$ and $v^0_4$ can be solved numerically. It is found that, in the regime where the perturbative RG method is applicable, the small perturbations due to $V_2$ and $V_4$ do not affect the renormalization of $V_1$ and $V_3$, whose flow show quite similar trends as indicated in Fig.2, i.e.,  $v_1$ and $v_3$ couplings are always dressed  to large positive and negative values, respectively. Meanwhile, $v_2$ and $v_4$ are renormalized to negligible values compared to $v_1$ and $v_3$. Therefore, it suggests that as long as the bare value of $v_2$ and $v_4$ is not too large, their renormalization do not have any qualitative consequence on the above conclusions, i.e., the same mean-field Hamiltonian can be constructed in certain low energy window and therefore the hybrid topological insulator will take place when the Kondo-coupling $g$ is larger than $g_c$. When the values of the bare interaction $v^0_2$ and $v^0_4$ are larger so that they are comparable to $v^0_1$ and $v^0_3$, their renormalization flows do not lead to negligible values. In this case, we have checked that $v_2$ also flows toward large negative value in the strong-coupling regime. As such, instead of the hybrid TI, the nematic phases in Table \uppercase\expandafter{\romannumeral1} would be the possible leading instability. We will leave the detailed discussion on this topic to further investigation.

\section{a summary}
We have revisited the traditional Kondo problem in the background of the recently discovered QBCP in two dimensional electron system. It is found that the Kondo problem, when combined with strongly-correlated effect, shows completely different physical behaviors compared to the traditional ones. On one hand, it is found that the Kondo-singlet will be destroyed when $g$ is less than the threshold value $g_c$, while robust for $g>g_c$. On the other hand, when $g>g_c$, the interplay of the two marginally relevant operators, \textit{i.e.}, the electron-electron interactions and the Kondo coupling leads to a coexistence state of the Kondo-singlet, QAH and the QSH state. This coexistence state has both nonzero Chern number and spin Chern  number, which is a new hybrid topological insulator since it is gapped in the bulk but enjoys a quantized charge Hall current and net transport of spin on its edges.
\begin{acknowledgments}
We wish to acknowledge Jiangtao Yuan, Haijun Zhang, Gaoming Tang, Yiming Pan, Hongyan Lu and Huaiqiang Wang for fruitful discussion. This work was supported by 973 Program under Grant No. 2011CB922103, and by
NSFC (Grants No. 60825402, No. 11023002, No. 91021003 and No. 11374246).
\end{acknowledgments}




\begin{thebibliography}{99}
\bibitem{rui} Since the quadratic perturbations are also renormalized under RG, the chemical potential, in priciple, needs to be fine-tuned to keep the electron density fixed. However, due to the divengence of the susceptibilities, gaps are opened at certain energy scale, which renders the fine-tune of the chemical potential unnecessary \cite{Song,Shankar}.
\end{thebibliography}
\end{document}